\documentclass[pra,twocolumn,superscriptaddress]{revtex4}
\usepackage{amsmath,amsfonts}

\def\bra#1{\mathinner{\langle{#1}|}}
\def\ket#1{\mathinner{|{#1}\rangle}}
\def\mn#1{\langle #1 \rangle}
\def\prjct#1{\mathinner{|{#1}\rangle}\!\!\mathinner{\langle{#1}|}}

\newcommand{\braket}[2]{\langle #1  |#2\rangle}

\def\text#1{\textrm{#1}}

\allowdisplaybreaks[1]

\begin{document}

\title{Bell Inequality Tests with Macroscopic Entangled States of Light}

\author{M. Stobi\'nska}
\affiliation{Max Planck Institute for the Science of Light, Erlangen, Germany}
\affiliation{Institute for Theoretical Physics II,
  Erlangen-N\"urnberg University, Erlangen, Germany}
\email[Corresponding author: M. Stobi\'nska,\hfil\break]{magda.stobinska@mpl.mpg.de}

\author{P. Sekatski}
\affiliation{Group of Applied Physics, University of Geneva, Geneva,
  Switzerland}

\author{A. Buraczewski}
\affiliation{Faculty of Electronics and Information Technology,
  Warsaw University of Technology, Warsaw, Poland}

\author{N. Gisin}
\affiliation{Group of Applied Physics, University of Geneva, Geneva,
  Switzerland}

\author{G. Leuchs}
\affiliation{Max Planck Institute for the Science of Light, Erlangen, Germany}
\affiliation{Institute for Optics, Information and Photonics,
  Erlangen-N\"urnberg University, Erlangen, Germany}

\begin{abstract}

  Quantum correlations may violate the Bell inequalities. Most of the experimental schemes
  confirming this prediction have been realized in all-optical Bell tests suffering from
  the detection loophole.  Experiment which closes this loophole and the locality loophole 
  simultaneously is highly desirable and remains challenging. A novel approach to a loophole-free 
  Bell tests is based on amplification of the entangled photons, i.e.\@ on macroscopic entanglement, 
  which optical signal should be easy to detect. However, the macroscopic states are partially 
  indistinguishable by the classical detectors. An interesting idea to overcome these limitations 
  is to replace the postselection by an appropriate preselection immediately after the amplification. 
  This is in the spirit of state preprocessing revealing hidden nonlocality. Here, we 
  examine one of possible preselections, but the presented tools can be used for analysis of 
  other schemes. Filtering methods making the macroscopic entanglement useful for Bell test and 
  quantum protocols are the subject of an intensive study in the field nowadays.

\end{abstract}

\maketitle

\section{Introduction}

Correlations between measurement results on entangled states are fascinating because they demonstrate sharply the difference between the classical and quantum description of the world. It is manifested in Bell inequality violation. Most of the experimental schemes confirming this prediction have been realized using photons. However, photons easily get lost and single-photon detectors are inefficient. Thus, all optical Bell tests postselect the events in which both photons are detected and suffer from the detection loophole. Experiment which closes this loophole and the locality loophole simultaneously is highly desirable and remains challenging. 

An appealing idea for a loophole-free Bell test is based on amplification of the entangled photons~\cite{DeMartini-PRL,DeMartini-single,Masha} by a unitary quantum cloner. If the amplification gain is high, the state is macroscopically populated and it should be easy to detect the optical signal. However, since the quantum macroscopic states are not fully distinguishable by the classical detectors, the measurement results will be affected by errors~\cite{DeMartini-PRL,Vitelli2010,Spagnolo2010,Sekatski2009}. It seems that postselection is unavoidable in Bell tests within the current technology.

An amplified single-photon is also fascinating as a potential macroscopic qubit candidate. Since the ideal amplification process is unitary, it preserves the entanglement originally present in the biphoton. Revealing such a micro-macro entanglement requires photon-number parity counts, again an unrealistic measurement with today's or near future technology.

An interesting idea to overcome these limitations is to replace the postselection by an appropriate preselection immediately after the amplification, before the measurement basis is chosen~\cite{Stobinska09}. It is inspired by the idea of hidden nonlocality~\cite{Popescu1995}. A mixed state, which is clearly nonlocal but does not violate any standard Bell inequality, can be preprocessed by a POVM measurement giving an access to quantum correlations hidden in its subspace which do violate CHSH inequality. The case of amplified entangled photons is similar. The classical detection mixes the quantum state and makes violation of CHSH inequality impossible. Thus, to be useful for any quantum protocol, these states need to be coherently filtered and preselection schemes are the subject of an intensive study now~\cite{Vitteli2010-2}. Intuitively, one taps a bit of light and keeps only quantum states that lead with high probability to clear measurement results. Here, we analyze one possible filtering method, which does not reveal Bell inequality violation, but increases the generation efficiency~\cite{Vitteli2010-2}. It paves the way towards other preselection schemes, necessary for performing Bell test or any quantum protocol. Mathematical tools developed here can be used for analysis of other experimental schemes.

This paper is organized as follows. We discuss the properties of macroscopic singlets in section \ref{macro}. In section \ref{TP} we present the analysis of preselection scheme.  We finish the paper with our conclusions.

\section{Macroscopic Entangled States}\label{macro}

Multiphoton quantum states of light are produced by phase covariant quantum cloners in phase sensitive coherent parametric amplification~\cite{DeMartini-PRL}. This method requires first a pair of linearly polarized photons created in a singlet state. The equatorial states of the Poincar\'e sphere of all polarization states are given by $a_{\varphi}^{\dagger} = \frac{1}{\sqrt{2}}(e^{i\varphi} a_H^{\dagger} + e^{-i\varphi} a_V^{\dagger})$, $a_{\varphi\perp}^{\dagger} = \frac{i}{\sqrt{2}}(e^{i\varphi} a_H^{\dagger} - e^{-i\varphi} a_V^{\dagger})$, where $a_{\varphi}^{\dagger}$ and $a_{\varphi\perp}^{\dagger}$ are creators for two orthogonal polarizations $\varphi$ and $\varphi\perp$. This subspace, parametrized by the polar angle $\varphi \in \langle 0,2\pi)$, is privileged for the phase covariant cloners, since here their Hamiltonian is rotationally invariant. We express the singlet state in this basis
\begin{equation}
  \ket{\psi^-}= \frac{1}{\sqrt{2}}\left(a^\dag_{\varphi }b^\dag_{\varphi \perp}-a^\dag_{\varphi \perp}b^\dag_{\varphi }\right)\ket{0}.
  \nonumber
\end{equation}
Next, one of its spatial modes is amplified to create a multiphoton state by passing the appropriate photon through a high gain $g$ nonlinear medium. This unitary evolution leads to the ``micro-macro'' singlet state
\begin{equation}
  \ket{\Psi^-} = C_g^{-2}\exp\left(\frac{T_g}{2}\left({a^\dag_{\varphi }}^2 +{a^\dag_{\varphi\perp }}^2 \right)\right)\ket{\psi^-},
  \nonumber
\label{micro-macro-operator}
\end{equation}
where $g$ is the amplification gain and $C_g = \cosh g$, $T_g=\tanh g$. For the purpose of further analysis, we rewrite it as a superposition of cuts with fixed photon numbers by expanding the exponent into its Taylor series
\begin{eqnarray}
\label{state repr1}
\ket{\Psi^-}&=&\sum_{N=0}^{\infty} \beta_N \ket{\psi^N}, \;
\ket{\psi^N} = \frac{1}{M} \left({a^\dag_{\varphi }}^2 + {a^\dag_{\varphi\perp }}^2 \right)^N \ket{\psi^-}, 
\nonumber \\
\ket{\psi^N} &=& \frac{1}{\sqrt{2}}( \ket{\psi_\varphi^N}_a\ket{1_{\varphi\perp}}_b -
  \ket{\psi_{\varphi\perp}^N}_a\ket{1_{\varphi}}_b ),
  \nonumber
\end{eqnarray}
where $\beta_N= C_g^{-2} T_g^N \sqrt{N+1}$, $M = 2^N \sqrt{N!(N+1)!}$ ensures normalization of $\ket{\psi^N}$, $\ket{\psi_{\varphi (\perp)}^N} =\frac{1}{M}\left({a^\dag_{\varphi }}^2 +{a^\dag_{\varphi\perp }}^2 \right)^N\ket{1_{\varphi(\perp)}}$. The components corresponding to the cut with $2N+1$ photons can be expressed directly in the Fock basis
\begin{equation}
\begin{aligned}
\kern-2em\ket{\psi_\varphi^N} =& \frac{1}{M}\!\sum_{k= 0}^N C^N_k \sqrt{(2k+1)!(2N-2k)!}\ket{2k+1,2N-2k},\kern-2em\\
\kern-2em\ket{\psi_{\varphi \perp}^N} =& \frac{1}{M}\!\sum_{k= 0}^N C^N_k\sqrt{(2k)!(2N-2k+1)!}\ket{2k,2N-2k+1},\kern-2em
\end{aligned}
\label{eq:macroqubits}
\end{equation}
where $C^N_k = \binom{N}{k}$. In experiment $\ket{\Psi^-}$ contains $4m \simeq 10^{4}$ photons on average, where $m=\sinh^2g$. The states in Eq.~(\ref{eq:macroqubits}) are orthogonal, but in high photon number regime detection is not single photon resolving~\cite{Masha} and they reveal effective overlap. This is an important issue for Bell inequality violation and preselection can solve it. 

In~\cite{Stobinska09} preselection is theoretically described by a projective measurement
\begin{equation}
  \mathcal{P}^{(Th)}_{N_{th}} = \sum_{k,l; k+l\ge N_{th}} |k,l\rangle \langle k,l|.
  \label{theoretical-preselection}
\end{equation}
It cuts off the low Fock photon number contributions below a threshold $N_{th}$ in the initial superposition where the overlap seems to be the largest. 
The preselected macro-states contain $N_{th}$ photons at least distributed over two polarization modes. We will refer to this operation as to the theoretical preselection. In the experiment it is approximated by a scheme consisting of an unbiased beamsplitter (BS) and a POVM measurement, given also by Eq.~(\ref{theoretical-preselection}) but for a different threshold $K_{th}$, which examines the intensity of the reflected beam and depending on the result rejects or passes the transmitted beam to the Bell test. We will call this scheme the beamsplitter preselection. It is described by 
$\mathcal{P}^{(BS)}_{K_{th}} = \mathrm{Tr}_r \circ \mathcal{P}^{(Th)}_{K_{th}} \circ \mathcal{U}_{BS}$, where $\mathcal{U}_{BS}$ is the beamsplitter action, $\mathrm{Tr}_r$ is trace operation over the reflected mode. We believe here that the theoretical and beamsplitter preselection operators converge pointwise for some set of parameters $\mathcal{P}^{(BS)}_{K_{th}}\left(\Psi^{(g)}\right) \to \mathcal{P}^{(Th)}_{N_{th}}\left(\Psi^{(g)}\right)$, where $\Psi^{(g)} = \Psi^- $ is given for the gain value $g$.

\section{Preselection} \label{TP}

Theoretical preselection modifies $\ket{\Psi^-}$ as follows
\begin{equation}
\ket{\Psi^-_{N_{th}}} = \sum_{N= N_{th}}^{\infty} \bar \beta_N \ket{\psi^N},
\nonumber
\end{equation}
where $\bar \beta_N$ is renormalized probability amplitude.

We consider a general diagonal measurement operator $\mathcal{O}$ acting locally on the modes of a ``micro-macro'' singlet
\begin{equation}
  \mathcal{O} = \sum_{k,l}\alpha_{kl} \prjct{k,l}\otimes \mathcal{O}_b,
  \label{observable}
\end{equation}
where $\ket{k,l}$ is an arbitrary basis for the Hilbert space of the macroscopic states and $\mathcal{O}_b$ is a Hermitian operator acting on the single
photon Hilbert space. The expectation value of $\mathcal{O}$ for $\ket{\Psi^-_{N_{th}}}$ equals
\begin{equation}
  \label{expectation-value}
  \bra{\Psi^-_{N_{th}}}\mathcal{O}
  \ket{\Psi^-_{N_{th}}}= \sum_{N,M= N_{th}}^{\infty} \bar \beta_N \bar \beta_M
  \bra{\psi^N} \mathcal{O}\ket{\psi^M}.
\end{equation}
The cross terms with $ N\ne M$ are zero for the following reasons. Let us consider a general term $\langle \psi^N\! \prjct{k,l}\otimes \mathcal{O}_b\ket{\psi^M}$. It is nonzero if and only if simultaneously $\braket{\psi^N}{k,l} \ne 0$ and $\braket{k,l}{\psi^M} \ne 0$. This is possible only if the states contain the same number of photons, i.e.\@ $2N+1= k+l$ and $k+l = 2M+1$, leading to $M=N$. It brings Eq.~(\ref{expectation-value}) to the diagonal form. If the Bell operator $\mathcal{B}$ is a sum of operators satisfying Eq.~(\ref{observable}), its mean value equals the convex sum
\begin{equation}
  \bra{\Psi^-_{N_{th}}}\mathcal{B}\ket{\Psi^-_{N_{th}}} = \sum_{N=
    N_{th}}^{\infty} \bar \beta_N^2 \bra{\psi^N} \mathcal{B}
  \ket{\psi^N}. \nonumber
\end{equation}
If this expression violates the local bound set by the CHSH inequality $\bra{\Psi^-_{N_{th}}}\mathcal{B}\ket{\Psi^-_{N_{th}}}>2$, at least one term $\bra{\psi^N} \mathcal{B} \ket{\psi^N}$ violates it. Therefore, we will consider each $\ket{\psi^N}$ separately.

The Bell operator for the CHSH inequality equals
\begin{align}\label{Bell}
\mathcal{B}=\mathcal{O}(\varphi_a)\otimes\mathcal{O}(\varphi_b)+
\mathcal{O}(\varphi_a)\otimes\mathcal{O}(\varphi_{b'})
\nonumber\\
+\mathcal{O}(\varphi_{a'})\otimes\mathcal{O}(\varphi_b)
-\mathcal{O}(\varphi_{a'})\otimes\mathcal{O}(\varphi_{b'}),
\end{align}
where one observer, Alice, measures the macroscopic part of the singlet and the other, Bob, measures the microscopic part. We assume the ideal measurement operator $\mathcal{O}(\varphi_b)= \prjct{1_{\varphi_b}}- \prjct{1_{\varphi_{b\perp}}}$ for the Bob's side, while for the Alice's side we take the threshold detection operator~\cite{Stobinska09} dictated by the expected modification of the state in the photon number space
\begin{equation}
  \label{projectors}
  \mathcal{O}(\varphi_a)=
  \mathcal{P}_{n_{\varphi_a}\le N_\sigma}\otimes
  \mathcal{P}_{n_{\varphi_a^\perp}> N_\sigma}-
  \mathcal{P}_{n_{\varphi_a}> N_\sigma}\otimes
  \mathcal{P}_{n_{\varphi_a^\perp}\le N_\sigma},
\end{equation}
where $N_\sigma$ is the threshold value. It projects onto two subspaces in the photon number space: with at least and at most (at most and at least) $N_\sigma$ photons in polarization $\varphi_a$ and $\varphi_a^\perp$, respectively. The proper value of $N_\sigma$ optimizes the observable to reveal the maximum amount of quantum correlations in polarization during the Bell test performed with photon number measurements. Taking $N_\sigma < N_{th}$  results in loosing the correlations. The best correlations are observed for $N_\sigma$ approximately equal to half of the total mean photon number in the state. 

We restrict $\mathcal{O}(\varphi_a)$ to the subspace with fixed number of photons where $\ket{\psi^N}$ belongs, and denote it $\mathcal{O}^N\!(\varphi_a)$. Applying the rotation between two basis by a relative angle $(\varphi_a-\varphi_b)$, which for the basis vectors transform as
\begin{align}
  \ket{1_{\varphi_b}} &= \cos(\varphi_a-\varphi_b)\ket{1_{\varphi_a}} + \sin(\varphi_a-\varphi_b)\ket{1_{\varphi_a^\perp}},
  \nonumber\\
  \ket{1_{\varphi_b^\perp}} &= -\sin(\varphi_a-\varphi_b)\ket{1_{\varphi_a}} + \cos(\varphi_a-\varphi_b)\ket{1_{\varphi_a^\perp}},\nonumber
  \nonumber
\end{align}
and noting that $\bra{\psi^N_{\varphi}} \mathcal{O}^N\!(\varphi)\ket{\psi^N_{\varphi\perp}}=0$ we compute the correlation $\bra{\psi^N}
\mathcal{O}^N\!(\varphi_a)\otimes\mathcal{O}(\varphi_b)\ket{\psi^N}$ to be equal to
\begin{equation}
  \frac{\cos(2(\varphi_a-\varphi_b))}{2}\left(\bra{\psi_{\varphi
        \perp}^N}\mathcal{O}^N\!(\varphi)\ket{\psi_{\varphi \perp}^N} -
    \bra{\psi_{\varphi}^N}\mathcal{O}^N\!(\varphi)\ket{\psi_{\varphi}^N}\right).
      \nonumber
\end{equation}
Noting that $\bra{\psi_{\varphi \perp}^N}\mathcal{O}^N\!(\varphi)\ket{\psi_{\varphi\perp }^N}=-\bra{\psi_{\varphi  }^N}\mathcal{O}^N\!(\varphi)\ket{\psi_{\varphi }^N}$ we simplify the above expression to the textbook form of the CHSH inequality for a singlet state
\begin{equation}
  \label{Cos}
  \cos(2(\varphi_a-\varphi_b))\bra{\psi_{\varphi
      \perp}^N}\mathcal{O}^N\!(\varphi)\ket{\psi_{\varphi \perp}^N}.
\end{equation}
It is well known that this inequality is violated only if the distinguishability $v(N,N_\sigma)= \bra{\psi_{\varphi    \perp}^N}\mathcal{O}^N\!(\varphi)\ket{\psi_{\varphi \perp}^N}$ exceeds $1/\sqrt{2}$. For $\mathcal{O}^N\!(\varphi)$ in Eq.~(\ref{projectors}) it takes the form
\begin{align}
  \label{prjctvis}
  v = \bra{\psi_{\varphi \perp}^N}\mathcal{P}_{n_{\varphi}\le N_\sigma}\otimes \mathcal{P}_{n_{\varphi^\perp}> N_\sigma} \ket{\psi_{\varphi \perp}^N} \nonumber\\-
  \bra{\psi_{\varphi \perp}^N}\mathcal{P}_{n_{\varphi}> N_\sigma}\otimes \mathcal{P}_{n_{\varphi^\perp}\le N_\sigma}
  \ket{\psi_{\varphi \perp}^N}.
\end{align}
Using the Fock space decomposition in Eq.~(\ref{eq:macroqubits}) it can be rewritten as a difference of sums computed in different photon number
regions
\begin{equation}
  \label{Vis sums}
  v(N,N_\sigma)=\left(\sum_{k\in S^\sigma_+} - \sum_{k\in S^\sigma_-}\right)F_N(k),
\end{equation}
where $F_N(k)= \left(\frac{C_k^N}{M}\right)^2(2k)!(2N-2k+1)!$.  The summation regions correspond to the constraints on $n_{\varphi}$ and $n_{\varphi^\perp}$ given in Eq.~(\ref{prjctvis})
\begin{align}
  S_+^\sigma =& \{k: k\leq N, 2k\leq N_\sigma, 2N-2k+1>N_\sigma \}, 
     \nonumber\\
  S_-^\sigma =& \{k: k\leq N, 2k > N_\sigma, 2N-2k+1\leq N_\sigma\}.
     \nonumber
\end{align}
If $N_\sigma \leq N$ these definitions are equivalent to
\begin{align}
  \label{regions1}
  S_+^\sigma =& \{k| 0\leq k \leq N_\sigma/2 \}\\
  \label{regions2}
  S_-^\sigma =& \{k| N - (N_\sigma -1)/2 \leq k \leq N \}.
\end{align}
Moreover, if $N_\sigma = N + p$ with a strictly positive $p$, then the regions $S_+^\sigma$ and $S_-^\sigma$ satisfy Eqs.~(\ref{regions1}) and (\ref{regions2}) for a new threshold $N_\sigma = N-p \leq N$. Therefore, it is sufficient to consider the case of $N_\sigma \leq N$. Using the expression in Eq.~(\ref{Vis sums}) it is easy to show that $v(N,N_\sigma +1)>v(N,N_\sigma-1)$ for every $N_\sigma < N$. The maximal distinguishability equals either $v_{max}= v(N,N)$ or $v_{max}=v(N,N-1)$ for $N$ even or odd, respectively. We introduce the symbol $\left[\frac{N}{2}\right]$ which equals to $N/2$ for even and $N/2-1/2$ for odd $N$. The maximal distinguishability equals
\begin{equation}
  v_{max}(N) = \left(\sum_{k=0}^{\left[\frac{N}{2}\right]}-\sum_{k=N-\left[\frac{N}{2}\right]+\frac{1}{2}}^{N}\right)F_N(k),
  \nonumber
\end{equation}
which can be further rewritten to a form where an analytic solution is found
\begin{equation}
  \label{vismax}
  v_{max}(N) = F_N\left(\left[\frac{N}{2}\right]\right) \!+\! \sum_{k= 0}^{\left[\frac{N}{2}\right]-1}\left(F_N(k)-F_N(N-k)\right), \nonumber
\end{equation}
with $F_N(k)-F_N(N-k)= \left(\frac{C_k^N}{M}\right)^2 (2k)!(2N-2k)! (2N-4k)$. Performing the summation yields
\begin{align}
  v_{max}(N) =& (C_{\left[\frac{N}{2}\right]}^N 2^{-N})^2 (N+1)\hspace{10 pt}
                \text{for}\hspace{5 pt} N \hspace{5 pt}\text{even}, \nonumber\\
  v_{max}(N) =& (C_{\left[\frac{N}{2}\right]}^N 2^{-N})^2 (N+2)\hspace{10 pt}
                \text{for}\hspace{5 pt} N \hspace{5 pt}\text{odd}. \nonumber
\end{align}
From this expression one can easily see that $v_{max}(2n)=v_{max}(2n-1)$ and $v_{max}(2n)>v_{max}(2(n+1))$. 
The distinguishability is sufficient to violate CHSH inequality only for $N \in \{0,1,2\}$. Using the Stirling approximation we show that the asymptotic value (the limit of high photon number) for the distinguishability of the macroscopic states equals
\begin{equation}
  \lim_{N\to\infty} v_{max}(N) = \frac{2}{\pi} < \frac{1}{\sqrt{2}}.
\end{equation}
Therefore, the theoretical preselection does not decrease the effective overlap between the highly populated macroscopic states and thus, the Bell test for the ``micro-macro singlet'' in the discussed scheme is impossible.

Theoretical preselection is an unphysical operation, but it approximates well the beamsplitter one. BS turns the ``micro-macro'' singlet into a mixture of terms corresponding to different number $M$ of reflected photons
\begin{equation}
\label{BSPS}
\rho_t= N \sum_{M= K_{th}}^{\infty}
\text{Tr}_r \mathcal{L}^M(\prjct{\Psi^-}) = N \sum_{M= K_{th}}^{\infty} p_M \,
\rho_M, \nonumber
\end{equation}
where $N = 1/\sum_{M= K}^{\infty} p_M$, $p_M = \text{Tr}_{a,v}\, \mathcal{L}^M(\prjct{\Psi^-}) $ is probability of reflecting $M$ photons and $\rho_M =\text{Tr}_v\, \mathcal{L}^M(\prjct{\Psi^-})/p_M $ is the state conditioned on this event. Using BS operation for two independent polarizations $e^{t_\gamma(a_{\varphi} \, v^\dag + a_{\varphi^\perp} v_\perp^\dag)}$, developing it on each side of the projector and tracing out the $\{v,v_{\perp}\}$ vacuum modes we find the explicit form of the state $\rho_M$
\begin{eqnarray}
\rho_M &=& \frac{1}{p_M}\sum_{N} \beta_N^2 \,\text{Tr}_v\, \mathcal{L}^M
(\prjct{\psi^N}) = \sum_N p_M^N\, \rho_M^N,
\nonumber\\[-4pt]
\rho_M^N &=& \frac{1}{C^{2N+1}_M}\sum_{n=0}^M  G^M_n a^n_{\varphi} a_{\varphi^{\perp}}^{M-n}
           \prjct{\psi^N} a^{\dag\, n}_{\varphi} a_{\varphi^{\perp}}^{\dag\,M-n}, \nonumber
\end{eqnarray}
\vskip-4pt\noindent
where $G^M_n = \frac{1}{n! (M-n)!}$ and $\rho_M^N$ describes $\ket{\psi^N}$ after $M$ photons have been lost. If $\rho_M$ violates Bell inequality, at least one $\rho_M^N$ violates it too.  The Bell operator equals $\mn{\mathcal{B}}_{\rho_M^N} = \text{tr}\left\{\rho_M^N \mathcal{B}\right\}
=\bra{\psi^N}\bar{\mathcal{B}}\ket{\psi^N}$. Repeating the steps (\ref{Bell})-(\ref{Cos}) we express $\mn{\mathcal{B}}_{\rho_M^N}$ using new distinguishability $v = \bra{\psi_{\varphi \perp}^N}\bar{\mathcal{O}}^N\!(\varphi)\ket{\psi_{\varphi \perp}^N}$ with
\begin{equation}
  \bar{\mathcal{O}}^{N}=\frac{1}{C^{2N+1}_M}\sum_{n=0}^M G^M_n
  a^{\dag\, n}_{\varphi} a_{\varphi^{\perp}}^{\dag\,M-n} \,\mathcal{O}^{N-\frac{M}{2}}\, a^n_{\varphi} a_{\varphi^{\perp}}^{M-n}.
  \nonumber
\end{equation}
\vskip-4pt\noindent
The operator $\bar{\mathcal{O}}^N$ can be expressed as a convex sum
\begin{equation}
  \bar{\mathcal{O}}^N =
  \sum_{N'_\sigma = \min(N_\sigma, M)}^{M+N_\sigma} p(N'_\sigma)
  \mathcal{O}^{N}_{N'_\sigma},
  \nonumber
\end{equation}
where $\mathcal{O}^{N}_{N'_\sigma}$ is $\mathcal{O}^{N}$ with a new threshold $N'_\sigma$. Thus, we brought the analysis back to the theoretical preselection.
\vskip-8pt

\section{Conclusions}
\vskip-8pt

In this paper we emphasized the necessity of preprocessing of quantum macroscopic states of light generated by the optimal quantum cloners in presence of classical detection. It gives hope for a loophole-free inequality test and enables application of these states in quantum protocols. Preselection engineers the state and this is easily noticeable in the photon number space. The observable to be measured has to be chosen accordingly to this modification. The idea is based on trade off between the physical and effective overlap measured by inefficient detectors. Filtering decreases the effective overlap at the cost of increasing the physical one. Efficient tool for following this trend provides the photon number distribution. Here, we tested the easiest example of preselection: the photon number sum, which is basis independent. The initial expectation was that the distinguishability will increase for high population $N \ge N_{th}$. However, in that case distinguishability cannot be improved, but generation efficiency is reinforced instead. Since the computation is analytically involved, the correct strategy is to gain as much as possible from internal structure of the examined state. Instead of manipulating and taking into account the whole state, it is possible to perform operations on its building blocks separately. This structure, for each preselection, has to found separately. This qualitatively allows to infer the physical properties of these states, e.g.\@ possibility of Bell inequality violation, which is especially useful in the limit of high photon population. Mathematical methods developed here can be used for analysis of other experimental schemes. The alternative to preselection schemes is to engineer the source of macroscopic entanglement by seeding it with other than single photon inputs to obtain the desired distinguishability of output states, which seems to be even more demanding. 
\vskip-8pt

\acknowledgments

M. S. and A. B. thank R. W. Chhajlany, P. Horodecki, R. Horodecki and F. Sciarrino for discussions. Calculations were carried out at CI TASK in Gda\'nsk.  This work was partially supported by Ministry of Science and Higher Education Grant No.\@ 2619/B/H03/2010/38.

\end{document}